\documentclass{article}
\usepackage{graphicx}  
\usepackage{amsmath}   
\usepackage[compress]{cite}
\usepackage{amssymb}   
\usepackage{bm} 
\usepackage{dcolumn}
\usepackage{color}
\usepackage{mathrsfs}
\usepackage{amsfonts}
\usepackage{varioref}
\usepackage{textcomp}

\RequirePackage[colorlinks,citecolor=blue,urlcolor=magenta,linkcolor=blue]{hyperref}
\allowdisplaybreaks
\labelformat{section}{Section #1} 
\labelformat{subsection}{Section #1} 
\labelformat{subsubsection}{Section #1}
\labelformat{subsubsubsection}{Section #1}
\labelformat{equation}{Eq.~(#1)} 
\labelformat{figure}{Fig.~#1} 
\labelformat{subfigure}{Fig.~\thefigure#1} 
\labelformat{table}{Table~#1} 
\labelformat{appendix}{Appendix #1}
\newcounter{mnotecount}[section]

\renewcommand{\themnotecount}{\thesection.\arabic{mnotecount}}

\newcommand{\mnotex}[1]
{\protect{\stepcounter{mnotecount}}$^{\mbox{\footnotesize
$
\bullet$\themnotecount}}$ \marginpar{
\raggedright\small\em
$\!\!\!\!\!\!\,\bullet$\themnotecount: #1} }

\begin{document}

\title{\bf Strong cosmic censorship conjecture for a charged BTZ black hole}

\author{Chiranjeeb Singha\footnote{chiranjeeb.singha@saha.ac.in}$~^{1}$, Sumanta Chakraborty\footnote{tpsc@iacs.res.in}$~^{2}$ and Naresh Dadhich\footnote{nkd@iucaa.in}$~^{3}$\\
$^{1}$\small{Theory Division, Saha Institute of Nuclear Physics, Kolkata 700064, India}\\
$^{2}$\small{ School of Physical Sciences, Indian Association for the Cultivation of Science, Kolkata 70032, India}\\
$^{3}$\small{IUCAA, Post Bag 4, Ganeshkhind, Pune 411007}}
 


\maketitle
\begin{abstract}

The strong cosmic censorship conjecture, whose validation asserts the deterministic nature of general relativity, has been studied for charged BTZ black holes in three dimensional general relativity, as well as for Nth order pure Lovelock gravity in d=2N+1 spacetime dimensions. Through both analytical and numerical routes, we have computed the ratio of the imaginary part of the quasi-normal mode frequencies with the surface gravity at the Cauchy horizon. The lowest of which corresponds to the key parameter associated with violation of strong cosmic censorship conjecture. Our results demonstrate that this parameter is always less than the critical value $(1/2)$, thereby respecting the strong cosmic censorship conjecture. This is in complete contrast to the four or, higher dimensional black holes, as well as for rotating BTZ black hole, where the violation of strong cosmic censorship conjecture exists. Implications and possible connection with the stability of the photon orbits have been discussed. 

\end{abstract}
\section{Introduction} \label{introduction}

Spacetimes with Cauchy horizon result into loss of the deterministic nature of gravitational theories, since evolution of an initial data cannot uniquely predict the events located at the future of the Cauchy horizon. Such a loss of deterministic nature spells doom on any gravitational theories, since this leads to distrust on the field equations of the respective gravitational theories. For example, both the Reissner-Nordstr\"{o}m and Kerr spacetime, solutions of Einstein's field equations, inherit Cauchy horizon \cite{Chandrasekhar:15209, Poisson:2009pwt}. Thus it seems that the most successful theory describing gravity, namely general relativity, is not deterministic in nature. To save the day, the strong cosmic censorship conjecture was imposed, which states that spacetime cannot be extended beyond the Cauchy horizon with the square-integrable connection. Note that this version is due to Christodoulou \cite{Dafermos:2014cua, Christodoulou:2008nj}, since earlier versions of the strong cosmic censorship conjectures can be violated by the charged or rotating black holes in general relativity, \'{a} la the mass inflation scenario \cite{Poisson1990,Ori1991,Eric2004,Bhattacharjee2016}. Further computations assert that the Christodoulou's version is indeed respected by all the asymptotically flat black holes in general relativity with Cauchy horizon \cite{Dafermos:2014cua}. This is because, the exponential blue-shift towards the Cauchy horizon always dominates over the power law fall-off in the late time. 

However, it turns out that the assumption of asymptotic flatness is crucial for the strong cosmic censorship to hold. If the spacetime is asymptotically de Sitter, then the late time fall-off of the perturbation modes falling into the black hole will be exponentially small, which may dominate over the exponential rise through the blue-shift near the Cauchy horizon. This results into extension beyond the Cauchy horizon, violating the strong cosmic censorship conjecture \cite{PhysRevD.61.064016, Costa:2014aia, Christodoulou:2008nj}. In mathematical terms, the perturbation modes have an exponential decay, $\phi \sim \exp(-\omega_{\rm I}u)\phi_{0}$, where $\omega_{\rm I}$ is the imaginary part of the lowest lying perturbation modes. While, the exponential blueshift near the Cauchy horizon will result into, $|\phi_{\rm cauchy}|^{2}\sim \exp(\kappa_{-}u)|\phi|^{2}$, with $\kappa_{-}$ being the surface gravity of the Cauchy horizon. Thus, it is evident that, if the dimensionless ratio $(\omega_{\rm I}/\kappa_{-})<(1/2)$, the perturbation modes will diverge, respecting the strong cosmic censorship conjecture. However, for $(\omega_{\rm I}/\kappa_{-})>(1/2)$, the perturbation modes will be regular, leading to a violation of the strong cosmic censorship conjecture \cite{Cardoso:2017soq,Hintz:2015jkj,Dafermos:2017dbw}. 

The above results demonstrate that a violation of the strong cosmic censorship conjecture can happen for classical gravity in four dimensional asymptotically de Sitter (dS) spacetimes. In particular, near extremal Reissner-Nordstr\"{o}m-dS black hole violates the strong cosmic censorship conjecture, while the Kerr-dS black hole respects the same (for the case of Kerr-Newman-de Sitter black hole, see \cite{Casals:2020uxa}). Thus in four spacetime dimensions, the charged and the rotating black holes, which are asymptotically dS, have opposite behaviour, as long as the strong cosmic censorship conjecture is considered (however, see \cite{Hollands:2019whz,Dafermos:2018tha}). For asymptotically AdS spacetimes, on the other hand, the decay of the perturbation modes are slower, in particular Logarithmic \cite{Holzegel:2011uu,Holzegel:2013kna} and hence the exponential blueshift is bound to dominate, thereby respecting the strong cosmic censorship conjecture. This is due to the existence of stable circular null orbits around the asymptotically AdS black holes, which could trap the modes to slow down the decay. However, the situation is drastically different in three dimensions, where it has been recently shown that a rotating BTZ black hole indeed violates the strong cosmic censorship conjecture \cite{Dias:2019ery}. Moreover, the presence of higher dimension, or higher curvature theories of gravity, or, even different kind of scalar fields can have major impact on the strong cosmic censorship conjecture \cite{Rahman:2018oso,Rahman:2020guv,Mishra:2020jlw,Destounis:2019omd,Guo:2019tjy,Sang:2022frw} (for effects due to Dirac field, see \cite{Ge:2018vjq,Destounis:2018qnb,Rahman:2019uwf,Liu:2019rbq}). Finally, if one takes into account semiclassical contributions, then the quantum corrections become strong enough, such that the strong cosmic censorship conjecture is respected in four spacetime dimensions \cite{Hollands:2019whz} (see also \cite{Bhattacharjee:2020gbo,Gwak:2018rba,Gim:2019rkl}). Again in three spacetime dimensions, the quantum correction identically vanishes, therefore the violation of the strong cosmic censorship conjecture at the classical level carries over to the semi-classical domain as well \cite{Dias:2019ery,Hollands:2019whz}.
 
So far, the analysis of the strong cosmic censorship conjecture has been carried out for rotating BTZ black holes \cite{Dias:2019ery}, where it was demonstrated that under linear perturbation, the Cauchy horizon is non-singular. However, as demonstrated in \cite{Emparan:2020rnp}, it follows that if one includes higher order perturbations and consider their backreaction on the Cauchy horizon of rotating BTZ black hole, the Cauchy horizon will turn out to be singular. This was pointed out by using the holographic duality between three dimensional rotating BTZ black hole with four dimensional Kerr-type black hole, which respect the strong cosmic censorship conjecture. Thus the inclusion of higher order perturbations make the rotating BTZ black hole respect the strong cosmic censorship conjecture. In this article, we want to explore whether the strong cosmic censorship conjecture holds for a charged BTZ black hole, within the realm of linear perturbation theory, both in general relativity and beyond. This also brings to the discussion the connection between stability of photon orbits and of the strong cosmic censorship conjecture. In particular, we also wish to study analogous situations in the context of pure Lovelock theories in the spirit of \cite{PhysRevD.100.084011,Dadhich2015StaticPL}. For generality we will also study the case of a (charged+rotating) BTZ black hole.

This article is arranged as follows --- In \ref{stability}, we study the stability of the photon circular orbits for a charged as well as (charged+rotating) BTZ black hole in general relativity for a preliminary indication towards the validity of strong cosmic censorship. To bolster our claim, in \ref{charged BTZ}, we present a numerical analysis for obtaining the perturbation frequencies for the charged BTZ black holes and demonstrate the status of the strong cosmic censorship conjecture once and for all. In \ref{Lovelock}, we present a detailed analysis both analytically as well as numerically, for charged black holes in pure Lovelock theories. Finally, we end with a brief discussion over the results obtained\footnote{On the notational side, we have used $c=G=1$ throughout the paper.}.

\section{Photon circular orbits for BTZ black holes}\label{stability}

In this section, we wish to study the photon circular orbits for a BTZ black hole, in order to understand their stability --- the first pointer towards the nature of strong cosmic censorship conjecture. We start with the charged BTZ black hole, whose metric is given by \cite{Hendi:2010px, Hendi:2020yah, Hendi:2015wxa, Tang:2016vmu, Martinez:1999qi},
\begin{equation}\label{eq01}
d s^2 = -f(r)d t^2+ f(r)^{-1} d r^2+ r^2 d \phi^2~;\qquad 
f(r)=-M+\frac{r^2}{L^2}- 2 q^2 \ln\left(\frac{r}{L}\right)~.
\end{equation}
Here $M$ is the mass of the black hole, which is a dimensionless quantity, $q$ represents the electric charge of the black hole, and the cosmological constant $\Lambda\equiv -(1/L^2)$, where $L$ is the AdS radius. Note that, asymptotically the Ricci scalar, as well as the Kretschmann scalar behaves as if the spacetime is asymptotically AdS, since the radial dependence on the metric due to the electric charge is only Logarithmic. Therefore, it is apparent that in the absence of the cosmological constant, i.e., with $\Lambda \rightarrow 0$, the above metric becomes asymptotically flat. It is worth pointing out that without a negative $\Lambda$, the spacetime would not inherit a horizon and hence no black holes would exist. Thus the requirement of a black hole spacetime is the existence of a negative $\Lambda$. The limit $\Lambda \rightarrow 0$ is subtle, since the AdS length scale appears in the Logarithm as well, resulting into divergent metric elements. Thus finiteness of metric demands $q$ to vanish as well, leading to flat spacetime throughout, consistent with non-dynamical nature of gravity in three dimensions.  

While determining the photon circular orbits \cite{Chakraborty:2021dmu,Mishra:2019trb}, first of all, we note that the metric is independent of the coordinates $t$ and $\phi$, therefore we have two constants of motion --- (a) the energy $p_{t} = -E$ and (b) the angular momentum $p_{\phi} = \mathbb{L}$. Therefore, the only non-trivial geodesic equation to solve is the radial one, which becomes,
\begin{equation}
 \dot{r}^2=\left[E^2-f(r)\frac{\mathbb{L}^2}{r^2}\right]\equiv \left[E^2-V_{\rm eff}(r)\right]~,
\end{equation}
where `dot’ denotes derivative with respect to the affine parameter along the null geodesic and $V_{\rm eff}(r)\equiv f(r)(\mathbb{L}^2/r^2)$ is the effective potential. Given the potential, one can immediately determine the location of the circular photon orbit $r_{\rm ph}$ by setting $V^{'}_{\rm eff} (r)$ to zero, yielding $2f=rf'$, where `prime' denotes derivative with respect to $r$. Computing $V_{\rm eff}'$ from the previous expression for $V_{\rm eff}$, one can get the radius of the only photon sphere for a charged BTZ black hole as,
\begin{eqnarray}\label{eq4}
r_{\rm ph}= L~\exp\left[\frac{1}{2}-\frac{M}{2 q^2}\right]~.
\end{eqnarray}
Note that in the limit $q\rightarrow 0$, the photon sphere vanishes identically. On the other hand, as the cosmological constant $\Lambda$ vanishes, the photon sphere tends to infinity, again an expected result. 

The stability of the above circular photon orbit depends on the sign of $V^{''}_{\rm eff}$, evaluated at $r_{\rm ph}$ \cite{Berry:2020ntz, Tang:2017enb}, whose explicit expression for charged BTZ black hole is given by,
\begin{eqnarray}
 V^{''}_{\rm eff}(r)\Big|_{r_{\rm ph}}
 = \frac{4 \mathbb{L}^2 q^2}{L^4}\exp\left[-2+2\frac{M}{q^2}\right]~,
\end{eqnarray}
which is always positive, as $\mathbb{L}$, $M$, $q$ are all positive and with negative $\Lambda$, $L^4$ is also positive. Thus it follows that stable photon orbits exist for the charged BTZ black hole and hence is a pointer that strong cosmic censorship conjecture is possibly respected for a charged BTZ black hole, which we study in detail in the next section through numerical analysis. 

Having discussed the case of the charged BTZ black hole, let us briefly consider the case of a rotating and charged BTZ black hole, whose metric is given by \cite{Banados:1992wn, Akbar:2007zz, AKBAR2007217},
\begin{equation}\label{eqbtzrot}
d s^2 = -g(r)d t^2+ g(r)^{-1} d r^2+ r^2  \left(d \phi-\frac{J}{2 r^2} dt\right)^2~;\qquad 
g(r)=-M+\frac{r^2}{L^2}+\frac{J^2}{4 r^2}- 2 q^2 \ln\left(\frac{r}{L}\right)~.
\end{equation}
Here $M$ is the mass of the black hole, again a dimensionless quantity, $J$ is the angular momentum, $q$ represents the electric charge of the black hole, and the cosmological constant $\Lambda=-1/L^2$, where $L$ is the AdS radius. Asymptotically, the Ricci and the Kretschmann scalar, behaves identically to that of the AdS spacetime. Such that in the absence of the cosmological constant, i.e., with $\Lambda \rightarrow 0$, the above metric becomes asymptotically flat. Again, finiteness of the metric in the limit of vanishing cosmological constant demands $q=0$. In which case it can be transformed to that of a flat spacetime. 

In this case as well, the metric components are independent of the $t$ and the $\phi$ coordinate, resulting into conserved energy $E$ and conserved angular momentum $\mathbb{L}$. The effective potential experienced by a massless particle in this charged and rotating BTZ black hole spacetime takes the form, 
\begin{align}
V_{\rm eff}(r)= \frac{\mathbb{L}^2}{r^2}\left[\frac{r^2}{L^2}-M+\frac{J}{b}- 2 q^2 \ln \left(\frac{r}{L}\right)\right]~,
\end{align}
where, $b=(\mathbb{L}/E)$ is the impact parameter of the massless particle. Computing, $V_{\rm eff}'$ and setting the same to zero, we obtain the radius of the photon sphere for a rotating charged BTZ black hole as,
\begin{eqnarray}\label{eq4rotbtz}
r_{\rm ph}= L~\exp\left[\frac{1}{2}+\frac{J}{2 b q^2}-\frac{M}{2 q^2}\right]~.
\end{eqnarray}
Note that, in the limit $q\rightarrow 0$, with $(J/M)>b$, the photon sphere diverges, while for $(J/M)<b$, the photon sphere identically vanishes. The only choice for which a non-trivial photon sphere exists correspond to $b=(J/M)$. As one can immediately verify, this is the choice for which circular photon orbits exists for a rotating BTZ black hole. However, with $q=0$ and $b=(J/M)$, the effective potential is a constant. This shows that the orbit is in the borderline of being stable.  

For checking the stability of photon orbits, one has to calculate the $V^{''}_{\rm eff}$ at the circular photon orbit $r_{\rm ph}$. The value of the same is given by,
\begin{eqnarray}
V^{''}_{\rm eff}(r)\Big|_{r_{\rm ph}}=\frac{4 \mathbb{L}^2 q^2}{L^4}\exp\left[-2- \frac{2 J}{b q^2}+2\frac{M}{q^2}\right]~,
\end{eqnarray}
which is always a positive definite quantity. This shows that even for the rotating charged BTZ black hole stable photon orbit do exists. However, for $q=0$, the $V''_{\rm eff}$ identically vanishes, as fit for the rotating BTZ black hole. Thus alike the charged BTZ black hole this demonstrates that one should naively expect the strong cosmic censorship conjecture to be respected for a rotating charged BTZ black hole as well, making the scenario with the rotating black hole a special one in nature.  

\section{Strong cosmic censorship conjecture for a charged BTZ black hole in general relativity}\label{charged BTZ}

Having derived the stability of the photon sphere as a pointer towards upholding the strong cosmic censorship conjecture, we would now like to seek direct evidence for the same by computing $\beta \equiv (\omega_{\rm I}/\kappa_{-})$ both analytically and numerically. The analytic computation hinges on the Lyapunov exponent of the photon sphere, while the numerical analysis will determine the lowest lying quasi-normal modes (QNMs) for perturbation of the charged BTZ black hole in general relativity.   

\subsection{Analytical estimation of the Lyapunov exponent and the strong cosmic censorship conjecture}

In this section, we would like to compute the analytical estimation of $\beta$ for a charged BTZ black hole. This requires computation of the imaginary part of the QNM through Lyapunov exponent and the surface gravity at the Cauchy horizon. In order to determine the location of the Cauchy horizon, let us try to solve the equation $f(r)=0$ from \ref{eq01}. From which, the horizon radii of a charged BTZ black hole are given by,
\begin{eqnarray}\label{chargedbtzhorizon}
r_{\mp}=L~\exp\left(-\frac{M}{2 q^2}- \frac{L_{\omega_{\pm}}}{2}\right)~,
\end{eqnarray}
where,
\begin{align}
L_{\omega_{+}}=\textrm{LambertW}_{0}\left[-\frac{1}{q^2} \exp\left(-\frac{M}{ q^2}\right)\right]~;
\qquad
L_{\omega_{-}}=\textrm{LambertW}_{-1}\left[-\frac{1}{q^2} \exp\left(-\frac{M}{ q^2}\right)\right]~.
\end{align}
Note that the solutions of the transcendental equation, $ye^{y}=x$ are, $y=\textrm{LambertW}_{0}(x)$ and $\textrm{LambertW}_{-1}(x)$, provided $-(1/e)\leq x\leq 0$. Here, $r_{+}$ and $r_{-}$ are the radius of the event and the Cauchy horizons, respectively. The condition for both the horizons to exist, i.e., with $r_{\pm}>0$, corresponds to, $M\geq q^2\left(1-\ln q^{2}\right)$. 

Having derived the locations of both the event and the Cauchy horizon, the computation of the surface gravity associated with the Cauchy horizon proceeds as follows,
\begin{align}
\kappa_{-}=\frac{1}{2}f'(r_{-})=\frac{r_{-}}{L^2}-\frac{q^2}{r_{-}}~,
\end{align}
where, the location of the Cauchy horizon $r_{-}$ is given by \ref{chargedbtzhorizon}, for a charged BTZ black hole. The Lyapunov exponent, on the other hand, is associated with the potential and its double derivative at the location of the photon sphere, which is given by \cite{Mishra:2020jlw,Rahman:2018oso, Cardoso:2008bp, Cardoso:2017soq},
\begin{eqnarray}\label{eq5}
\lambda=\sqrt{\frac{f(r_{\rm ph})}{2}\left(\frac{2 f(r_{\rm ph})}{r^2_{\rm ph}}-f''(r_{\rm ph})\right)}~.
\end{eqnarray}
From the expression for $f(r)$ from \ref{eq01} and the radius of the photon sphere from \ref{eq4}, one can get Lyapunov exponent for a charged BTZ black hole as,
\begin{equation} \label{eq6}
\lambda=\sqrt{\frac{f(r_{\rm ph})}{r^2_{\rm ph}}\Big[- M - q^2\left\{1+2 \ln\left(\frac{r_{\rm ph}}{L}\right)\right\}\Big]}~.
\end{equation}
Given the expressions for $\lambda$ and $\kappa_{-}$ from above, the expression for $\beta_{\rm ph}=(\lambda/2 \kappa_{-})$ \cite{Mishra:2020jlw, Rahman:2018oso,Cardoso:2008bp,Cardoso:2018nvb,Dias_2019,Cardoso:2017soq} for a charged BTZ black hole can be written as,
\begin{eqnarray}\label{eq7}
\beta_{\rm ph}=\frac{\sqrt{f\left(r_{\rm ph}\right)\Big[-M-q^2\left\{1+2\ln\left(\frac{r_{\rm ph}}{L}\right)\right\}\Big]}}{r_{\rm ph}^2\left(\frac{2 r_{-}}{L^2}- \frac{2 q^2}{r_{-}}\right)}~.
\end{eqnarray} 
Finally, using the value for $r_{\rm ph}$ from \ref{eq4}, the expression for $\beta_{\rm ph}$ takes the following form,
\begin{equation}\label{ph}
\beta_{\rm ph}
=\frac{q}{L}\frac{\sqrt{2\left(q^2-e^{1-M/q^2}\right)}}{\exp\left(\frac{1}{2}-\frac{M}{2q^{2}} \right)\left(\frac{2 r_{-}}{L^2}- \frac{2 q^2}{r_{-}}\right)}~.
\end{equation}
The above expression explicitly determines the quantity $\beta_{\rm ph}$ for the photon sphere modes of a charged BTZ black hole analytically. Note that this analysis is incomplete, since the Lyapunov exponent depicts the imaginary parts of the QNMs in the eikonal limit. Thus it is not clear if the imaginary parts of the QNMs associated with low angular momentum values are smaller than the one in the eikonal limit. In order to determine the same we need to numerically find out the QNM frequencies and then obtain $\beta$ to demonstrate the validity/violation of the strong cosmic censorship conjecture.
\subsection{Numerical estimation of the lowest lying QNMs and the strong cosmic censorship conjecture}

Having described the Lyapunov exponent and the associated expression for the dimensionless ratio $\beta$, let us try to understand the same from a numerical analysis of the QNMs. This is because, the analytical computations refer to the eikonal limit (i.e., QNMs with large angular momentum). Thus it may happen that the lowest lying QNMs are those with smaller values of angular momentum and then the analytical estimation of $\beta$, presented above, would be wrong. Keeping this caveat of the analytical computation in mind, we wish to study the QNMs with smaller values of the angular momentum $\ell$, through numerical methods. 
 
For this purpose, we start by considering the perturbation of the charged BTZ black hole by a massless scalar field $\Phi$, such that the evolution of the perturbation is governed by the Klein-Gordon equation $\Box  \Phi=0$. Given the periodicity of the metric in time and angular coordinate, it is useful to choose the following ansatz for the perturbing scalar field $\Phi$ \cite{Mishra:2020jlw},
\begin{equation} \label{eq12}
\Phi(t,r,\phi)=e^{-i \omega t}~\frac{R(r)}{\sqrt{r}}~e^{i \ell \phi}~,
\end{equation}
which leads to the following master equation for the radial part $R(r)$ of the scalar perturbation $\Phi$,
\begin{equation}\label{eq13}
\left[\frac{\partial ^2}{\partial r_{*}^2}+\omega^2-v_{\rm eff}(r)\right]R(r)=0~.
\end{equation}
Here, 
\begin{equation}
v_{\rm eff}(r)=f(r)\left(\frac{\ell^2}{r^2}-\frac{f(r)}{4 r^2}+\frac{f'(r)}{2r}\right)~,
\end{equation}
is the effective potential experienced by the radial perturbation $R(r)$. Note that in the above expression $f(r)$ is the metric function defined in \ref{eq01}, such that the Tortoise coordinate $r_{*}$ gets defined through the differential equation $(dr/d r_{*})=f(r)$. The nth overtone QNM frequency $\omega_{n}$ is defined as the eigenvalue of \ref{eq13}, that corresponds to an ingoing mode at the event horizon, $r_{+}$, and an outgoing mode which tends to zero near the infinity
\begin{eqnarray}\label{eq14}
R(r\to r_{+})\sim e^{-i \omega r_{\star}}~\textrm{ and}~R(r\to \infty)\sim 0~.
\end{eqnarray}
The vanishing of the perturbation modes at infinity is motivated by the observation that the potential $v_{\rm eff}(r)$ diverges at infinity, therefore we require that perturbation vanishes there. Only a discrete set of complex frequencies $\omega$, namely the QNM frequencies will satisfy the above boundary conditions. In particular, we will be interested in the lowest lying QNM. 

\begin{figure*}[h!]
\centering
\includegraphics[scale=0.6]{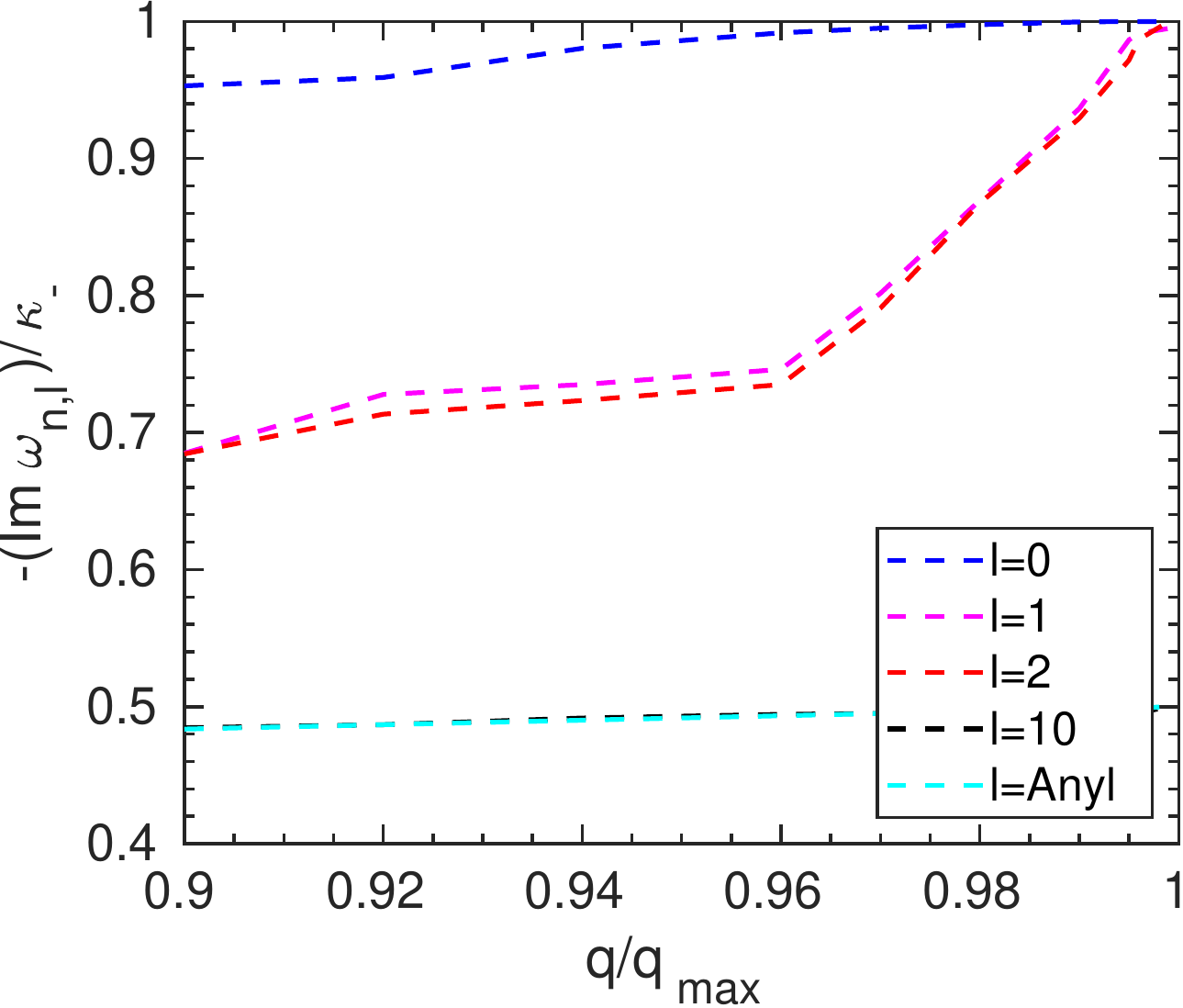}
\caption{Here we have plotted the ratio $\{-(\textrm{Im} \omega_{n,l})/ \kappa_{-}\}$ against $(q/q_{max})$ for different values of the angular momentum $\ell$.  The lowest lying curve depicts $\beta$ and as evident is less than the critical value (1/2). Here we have taken the values of the AdS length scale and the mass of the black hole to be $L=M=1$.}
\label{figure1} 
\end{figure*}

\begin{figure*}[h!]
\centering
\includegraphics[scale=0.6]{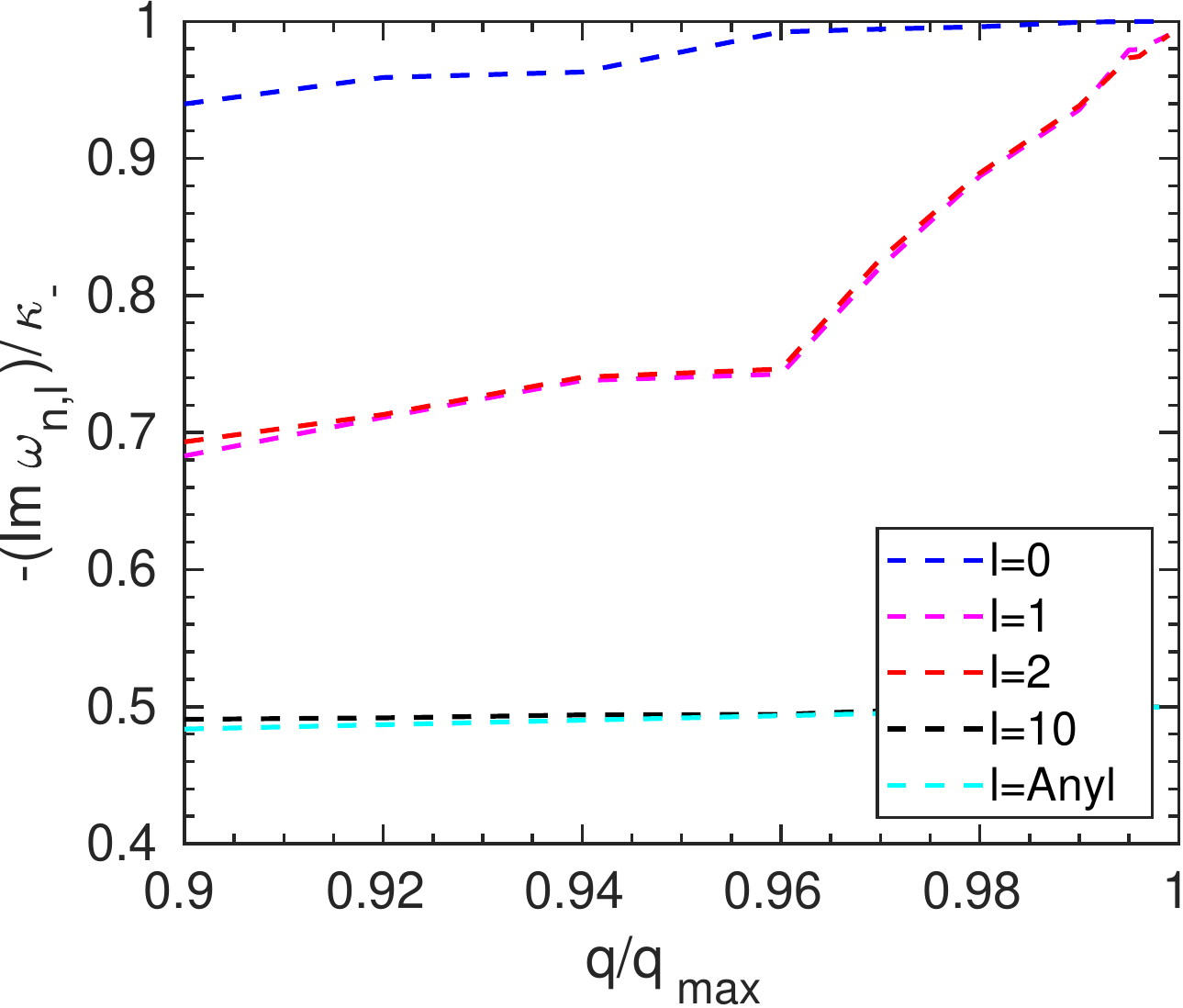}
\caption{In the above, we have depicted the ratio $\{-(\textrm{Im} \omega_{n,l})/ \kappa_{-}\}$ against $(q/q_{\rm max})$ for different choices of the angular momentum $\ell$ and for AdS radius and mass $L=2$ and $M=1$, respectively. The lowest lying curve corresponds to $\beta$ and is always less than the critical value $(1/2)$.}
\label{figure2} 
\end{figure*}

The numerical computation of the QNMs is performed using the procedure elaborated in \cite{Cardoso:2001bb,Berti:2009kk,Cardoso:2001hn}. Having obtained the QNMs using the symbolic manipulation package MATHEMATICA, we obtain the complex frequencies of the QNMs, which are the first ingredients that go into the definition of $\beta$. The computation of the surface gravity for the Cauchy horizon. i.e., $\kappa_{-}$ proceeds analytically, and, hence, the numerical estimation for $\{-(\textrm{Im}~\omega_{n,l})/ \kappa_{-}\}$ can be obtained, whose minimum value would yield the estimation for $\beta$. We calculate numerical values of $\{-(\textrm{Im}~\omega_{n,l})/ \kappa_{-}\}$ for the lowest-lying QNMs for different choices of angular momentum $\ell$ and the ratio $(q/q_{\rm max})$ in \ref{label1} and \ref{label2}, respectively. Also, we plot $\{-(\textrm{Im}~\omega_{n,l})/ \kappa_{-}\}$ versus the ratio $(q/q_{\rm max})$, where $q_{\rm max}$ corresponds to the extremal value of the electric charge, for different choices of $\ell$ in \ref{figure1} and \ref{figure2}. We further show that the ratio $\{-(\textrm{Im}~\omega_{n,l})/ \kappa_{-}\}$ for the lowest-lying QNM, i.e., $\beta$ is always less than half. Therefore, the strong cosmic censorship conjecture is respected for a charged BTZ black hole in comparison to the rotating one. 

Note that, even if there are other QNM modes present in our model, like the internal QNM modes for rotating BTZ black hole, they will be further sub-dominant and hence the strong cosmic censorship conjecture will hold stronger. However, the presence of the Logarithmic term in the metric coefficients results into a completely different QNM structure for charged BTZ black hole. Therefore the internal QNMs, which are present for rotating BTZ black hole and plays the pivotal role in violating the strong cosmic censorship conjecture, are absent for charged BTZ black holes. Thus the lowest lying QNMs computed in this work numerically are indeed the lowest lying, there are no other classes of QNMs present in our model.

So far the analysis is for general relativity. In what follows we will try to extend our analysis in the context of higher curvature gravity theories as well. This is of importance, since as demonstrated in \cite{Mishra:2020jlw}, presence of higher curvature terms seem to result into a stronger violation of the strong cosmic censorship conjecture. Thus one may ask, even if the strong cosmic censorship conjecture is respected in general relativity, they may get violated in a higher curvature scenario. In order to explore the same, we will discuss the case of a charged BTZ black holes in the context of Lovelock gravity.

\begin{table*}[h!]     
\centering
\begin{tabular}{cccccccccccccccccccccccccccccccccc}         
\hline\hline                
 $q/q_{max}$ && $\ell=0$ && $\ell=1$ && $\ell=2$ && $\ell=10$  && $\ell=10$ (Analytical)\\ \hline
 0.9 && 0.952944 && 0.684887 && 0.684413 && 0.484647 && 0.483657\\
 0.92 && 0.959046 && 0.727779 && 0.713391 && 0.486837 && 0.486867\\
 0.94 && 0.980311 && 0.734965 && 0.723364 && 0.491705 && 0.490109\\
 0.96 && 0.991626 && 0.74614 && 0.735 && 0.4944 && 0.493381  \\
 0.97 && 0.994995 && 0.801585 && 0.790740 && 0.495008 && 0.495026  \\
 0.98 && 0.997563 && 0.869417 && 0.867348 && 0.496041 && 0.496678  \\
 0.99 && 0.99957 && 0.936341 && 0.929226 && 0.497589 && 0.498336  \\
 0.995 && 0.999878 && 0.986484 && 0.971652 && 0.499504  && 0.499167  \\ 
 0.996 && 0.999826 && 0.991247 && 0.986762 && 0.495511  && 0.499334\\
 0.999  && 0.999883 && 0.994834 && 0.999835 && 0.499916  && 0.499833\\
\hline\hline                                                    
\end{tabular}                                
\caption{Numerical values of the ratio $\{-(\textrm{Im} \omega_{n,l})/ \kappa_{-}\}$ have been presented for the lowest lying QNMs for different choices of the angular momentum $\ell$ and the ratio $(q/q_{\rm max})$. Here we have considered the value of the AdS length scale and the mass of the black hole to be, $L=M=1$.}
\label{label1}             
\end{table*}   
\begin{table*}[h!]       
\centering
\begin{tabular}{cccccccccccccccccccccccccccccccccc}         
\hline\hline                
 $q/q_{max}$ && $\ell=0$ && $\ell=1$ && $\ell=2$ && $\ell=10$  && $\ell=10$ (Analytical)\\ \hline
 0.9 && 0.939713 && 0.682941 && 0.693184 && 0.490807 && 0.483657 \\
 0.92 && 0.959043 && 0.711081 && 0.713135 && 0.491822 && 0.486867 \\
 0.94 && 0.96299 && 0.737917 && 0.740409 && 0.494006 && 0.490109  \\
 0.96 && 0.992335 && 0.74261 && 0.746318 && 0.49439 && 0.493381  \\
 0.97 && 0.994333 && 0.820761 && 0.82662 && 0.496916 && 0.495026  \\
 0.98 && 0.996018 && 0.887026 && 0.889284 && 0.497062  && 0.496678  \\
 0.99 && 0.999342 && 0.935622 && 0.938388  && 0.497149  && 0.498336  \\
 0.995 && 0.999838 && 0.979173 && 0.973376 && 0.49951 && 0.499167  \\
 0.996 && 0.9999 && 0.979465 && 0.974342 && 0.499663 && 0.499334  \\
 0.999 && 0.99995 && 0.99 && 0.99 && 0.4998 && 0.499833  \\
\hline\hline                                                    
\end{tabular}                                
\caption{Numerical values of the ratio $\{-(\textrm{Im} \omega_{n,l})/ \kappa_{-}\}$ have been presented for the lowest lying QNMs for different choices of the angular momentum $\ell$ and the ratio $(q/q_{\rm max})$. Here we have considered the value of the AdS length scale and the mass to be, $L=2$ and $M=1$, respectively.}
\label{label2}             
\end{table*}


\section{Strong cosmic censorship conjecture for a charged BTZ black hole in pure Lovelock gravity}\label{Lovelock}

In this section, we wish to study the strong cosmic censorship for BTZ-like black holes in higher curvature gravity. The scenario in which such BTZ-like black holes appear naturally is in the paradigm of pure Lovelock gravity. Generally, Lovelock theories are special, since these are the theories in which despite the presence of higher curvature terms in the Lagrangian, the field equations are still second order. The Lovelock Lagrangian in a generic spacetime dimension $d$ contains $N$ number of terms, polynomial in the Riemann curvature tensor, such that $N\leq (d/2)$ \cite{PhysRevD.100.084011,Dadhich:2012cv,Dadhich2015StaticPL,Gannouji:2013eka,PhysRevD.93.064009}. Thus for a given Lovelock Lagrangian of order $N$, i.e., involving $N$ Riemann tensors, in spacetime dimensions $d=2N+1$, we will encounter behaviour of gravity to be similar to that of general relativity in three dimensions. This ensures that BTZ-like solutions will exist for pure Lovelock theories, while for general Lovelock theories, such solutions do not exist. Thus for our purpose, pure Lovelock theories play a crucial role.   

Here we consider a charged BTZ black hole in $N$th order pure Lovelock gravity in $d$ spacetime dimensions, such that, $d=2N+1$. This scenario is identical to that of the three dimensional general relativity, since general relativity is also a pure Lovelock theory with $N=1$. To start with we wish to note down that the metric for a charged black hole in $N$th order pure Lovelock gravity in $d$ spacetime dimensions (arbitrary $d$) is given by, \cite{Chakraborty:2020ifg,Chakraborty:2016qbw}, 
\begin{equation}\label{eq15}
ds^{2}=-f(r)dt^2+ f(r)^{-1} d r^2+ r^2 d \Omega_{d-2}^2~;\qquad
f(r)=1-\left(\Lambda r^{2N}+ \frac{2^{N} M}{r^{d-2N-1}}-\frac{q^2}{r^{2d-2N-4}}\right)^{1/N}~.
\end{equation}
Here $M$ is the mass of the black hole, $q$ represents the electric charge of the black hole and $\Lambda$ is the cosmological constant, which is negative. Due to the negative value of $\Lambda$, real solutions are possible only for odd values of the Lovelock order $N$, thus in what follows we consider only these values of $N$. Then the metric function $f(r)$ becomes, $f(r)=1-[\Lambda r^{2N}+ 2^{N} M-(q^2/r^{2N-2})]^{1/N}$. Note that this result is not valid for $N=1$, where the coefficient of the charge term is Logarithmic in the radial coordinate $r$. The location of the horizons are given by the solutions of the equation $f(r)=0$, which has two real roots, when $\Lambda$ is negative. As usual, the radius of the event horizon is denoted as $r_{+}$, and the radius of the Cauchy horizon is denoted by $r_{-}$.

\begin{figure*}[h!]
\centering
\includegraphics[scale=0.6]{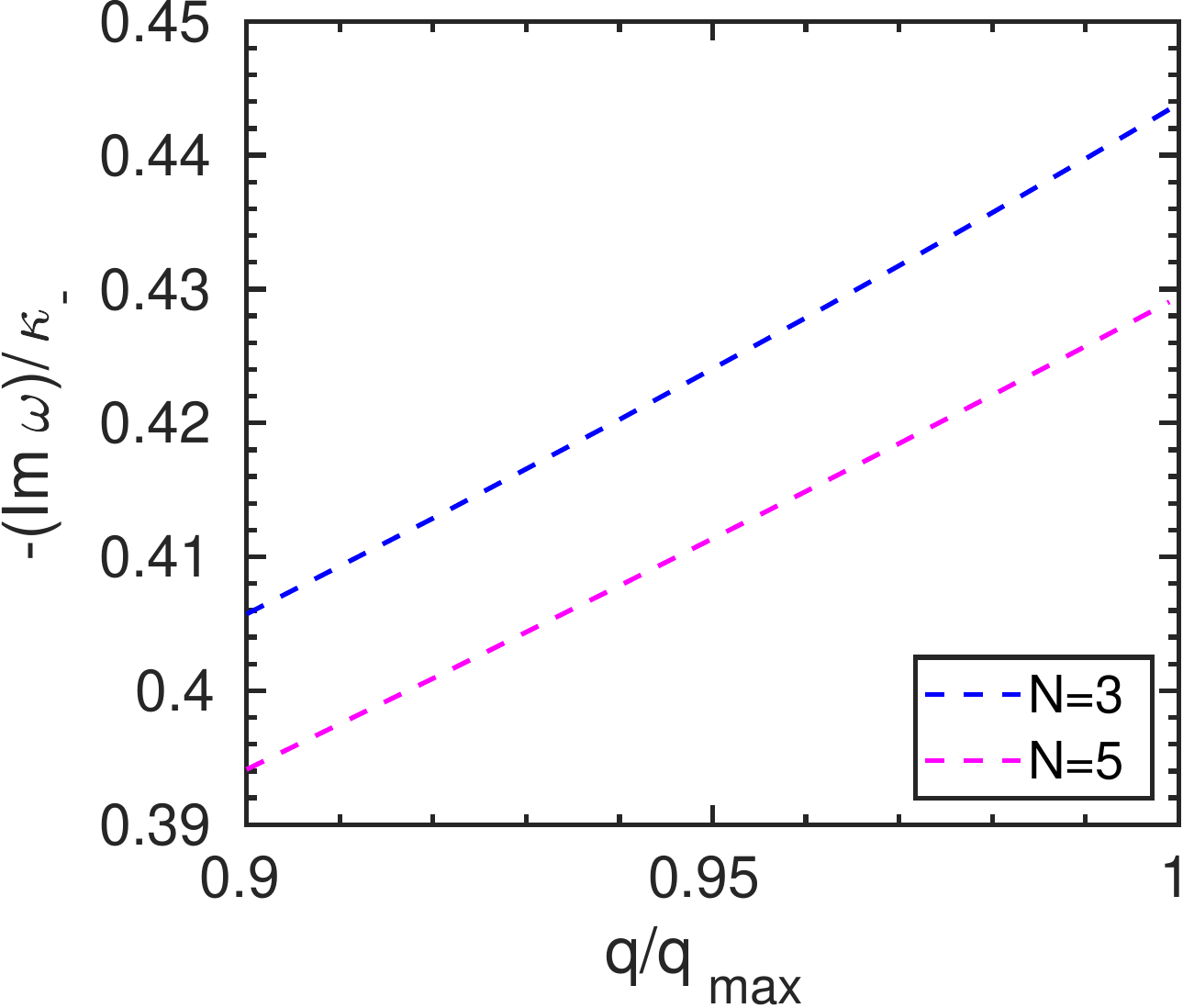}
\caption{We have plotted the analytical estimation of the ratio $-(\textrm{Im}~\omega/\kappa_{-})$ against the ratio $(q/q_{\rm max})$, where we have taken the value of the cosmological constant to be $\Lambda=-0.1$ and the mass to be $M=1$, for two different choices of the Lovelock order $N$.}
\label{figure03} 
\end{figure*}

\begin{figure*}[h!]
\centering
\includegraphics[scale=0.6]{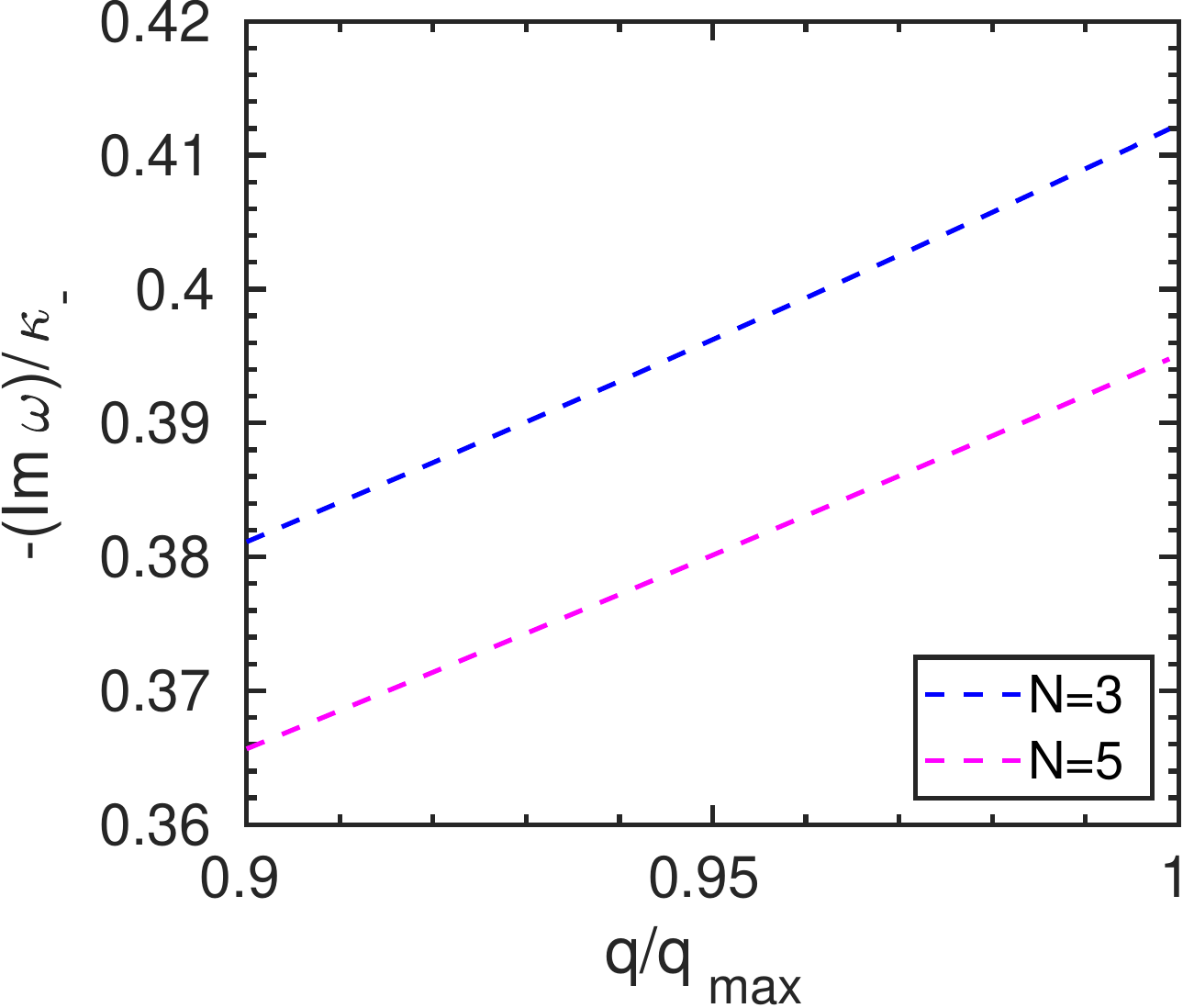}
\caption{The ratio $-(\textrm{Im}~\omega/\kappa_{-})$ has been plotted against $(q/q_{\rm max})$, where we have taken the value of the cosmological constant to be $\Lambda=-0.06$ and the black hole mass $M=1$, for two different choices of the Lovelock order $N$.}
\label{figure04} 
\end{figure*}

Due to the presence of the Cauchy horizon there is a possibility of the violation of the strong cosmic censorship conjecture, which we wish to explore further in this section. However, due to the involved nature of the metric element $f(r)$, we will not present any analytical expressions for the Lyapunov exponent or, of the surface gravity here, but both of them can be calculated using procedures elaborated above, as well as in \cite{Mishra:2020jlw, Rahman:2018oso, Cardoso:2008bp,Cardoso:2017soq,Cardoso:2018nvb, Dias_2019}. Therefore, it is possible to compute the ratio $-(\textrm{Im}\omega/\kappa_{-})$, using the Lyapunov exponent of the photon sphere located at $r_{\rm ph}$, determined by the condition $2f(r)=rf'(r)$, where ``prime'' represent the derivative with respect to $r$. The results of such an analytical computation have been presented in \ref{label03} and \ref{label04} for different choices of the cosmological constant $\Lambda$ and black hole mass $M$. We have also plotted the values of the ratio $-(\textrm{Im}\omega/\kappa_{-})$ against $(q/q_{\rm max})$, where $q_{\rm max}$ is the extremal value of the electric charge, for different choices of $N$ in \ref{figure03} and \ref{figure04}, respectively. As evident from \ref{label03} and \ref{label04}, as well as from the \ref{figure03} and \ref{figure04}, the higher the value of the Lovelock order $N$, strong cosmic censorship conjecture is respected more strongly. This is in complete contrast to the corresponding scenarios in higher dimensional spacetimes \cite{Rahman:2018oso, Liu:2019lon}. However, for completeness, in the next section, we will also numerically determine the QNM frequencies to ensure that the strong cosmic censorship conjecture is indeed respected for charged BTZ black holes in pure Lovelock theories of gravity.  
\begin{table*}[h!]            
\centering
\begin{tabular}{cccccccccccccccccccccccccccccccccc}         
\hline\hline                
 $q/q_{max}$ && $N$ && Analytical && $N$ && Analytical \\ \hline
 0.9 && 3 && 0.405728 && 5 && 0.394121\\
 0.92 && 3 &&  0.412883 && 5 && 0.40092\\
 0.94 && 3 &&   0.420254 && 5 && 0.407841\\
 0.96 && 3 &&  0.427856 && 5 && 0.41489\\
 0.97 && 3 && 0.43175 && 5 && 0.418466  \\
 0.98 && 3 && 0.435709 && 5 && 0.422078  \\
 0.99 && 3 && 0.439739 && 5 && 0.425727  \\
 0.995 && 3 && 0.441781 && 5 && 0.427566\\ 
 0.996 && 3 && 0.442191 && 5 && 0.427935\\
 0.999 && 3 && 0.443428 && 5 && 0.429045\\
\hline\hline                                                    
\end{tabular}                                
\caption{Analytical values of the ratio $-(\textrm{Im}~\omega/\kappa_{-})$ have been presented for two different values of the Lovelock order $N$. Here we have taken the cosmological constant and the mass to read, $\Lambda=-0.1$ and $M=1$, respectively.}
\label{label03}             
\end{table*}  
\begin{table*}[h!]            
\centering
\begin{tabular}{cccccccccccccccccccccccccccccccccc}         
\hline\hline                
 $q/q_{max}$ && $N$ && Analytical && $N$ && Analytical \\ \hline
 0.9 && 3 && 0.381119 && 5 && 0.365658  \\
 0.92 && 3 && 0.38704 && 5 && 0.371384  \\
 0.94 && 3 && 0.393112 && 5 && 0.377187  \\
 0.96 && 3 && 0.399341 && 5 && 0.38307  \\
 0.97 && 3 && 0.402517 && 5 && 0.386043  \\
 0.98 && 3 && 0.405734 && 5 && 0.389037  \\
 0.99 && 3  && 0.408994 && 5 && 0.392054  \\
 0.995 && 3 && 0.4106 && 5 && 0.393571  \\
 0.996 && 3 && 0.410971 && 5 && 0.393875  \\
 0.999 && 3 && 0.411965 && 5 && 0.394788 \\
\hline\hline                                                    
\end{tabular}                                
\caption{We have plotted the analytical values of the ratio $-(\textrm{Im}~\omega/\kappa_{-})$ for different value of the Lovelock order $N$ for the value of the cosmological constant $\Lambda=-0.06$ and mass $M=1$.}
\label{label04}             
\end{table*}

\subsection{Numerical estimation for the strong cosmic censorship conjecture for a charged BTZ black hole in pure Lovelock gravity}

The analytical computation presented above already provides a hint that strong cosmic censorship conjecture is respected for charged BTZ black holes in higher order pure Lovelock gravity more strongly, in comparison to general relativity. In this section, on the other hand, we will present a numerical evidence for the same. For this purpose, let us start by describing the dynamics of a massless scalar field $\Phi$ on a charged BTZ black hole in pure Lovelock gravity. The evolution of the perturbation is governed by the Klein-Gordon equation $\Box \Phi=0$.  We make the following ansatz for the field $\Phi$ \cite{Mishra:2020jlw},
\begin{equation}\label{eq16}
\Phi(t,r,\Omega)=\sum_{l,m} e^{-i \omega t}~\frac{\mathcal{R}(r)}{r^{(2N-1)/2}}~Y_{lm}(\Omega)~,
\end{equation}
which leads to the following master equation for the radial perturbation $\mathcal{R}(r)$,
\begin{equation}\label{eq17}
\left(\frac{\partial ^2}{\partial r_{\star}^2}+\omega^2-v_{\rm eff}(r)\right)\mathcal{R}(r)=0~,
\end{equation}
where, 
\begin{equation}
v_{\rm eff}(r)=f(r)\left(\frac{\ell(\ell+2N-2)}{r^2}+\frac{(2N-1)(2N-3)f(r)}{4 r^2}+\frac{(2N-1)f'(r)}{2r}\right)~,
\end{equation}
is the effective potential expressed above in terms of the metric function $f(r)$ and $dr_{\star}=\{dr/f(r)\}$ is the tortoise co-ordinate. The QNM frequency $\omega_{n}$ is defined as the eigenvalue of \ref{eq17} that corresponds to the following boundary conditions: ingoing modes at the event horizon, $r_{+}$, and the mode tends to zero near the infinity, 
\begin{eqnarray}\label{eq18}
\mathcal{R}(r\to r_{+})\sim e^{-i \omega r_{\star}}~\textrm{ and}~\mathcal{R}(r\to \infty)\sim 0~.
\end{eqnarray}
Now for computing the quasinormal modes numerically, we follow the procedure and use the MATHEMATICA package developed in \cite{Cardoso:2001bb, Berti:2009kk, Cardoso:2001hn}, but modified to meet our criterion. Defining the appropriate radial coordinates and imposing the relevant QNM boundary conditions at the event horizon and at infinity we obtain the complex frequencies of the QNMs, which are the first ingredients that go into the ratio $\{-(\textrm{Im}~\omega_{n,l})/ \kappa_{-}\}$ and the lowest of which determines $\beta$. The computation of surface gravity $\kappa_{-}$ for the Cauchy horizon can also be performed similarly, and, hence, the numerical estimation for $\{-(\textrm{Im}~\omega_{n,l})/ \kappa_{-}\}$ can be obtained.

\begin{figure*}[h!]
 \centering
\includegraphics[scale=0.6]{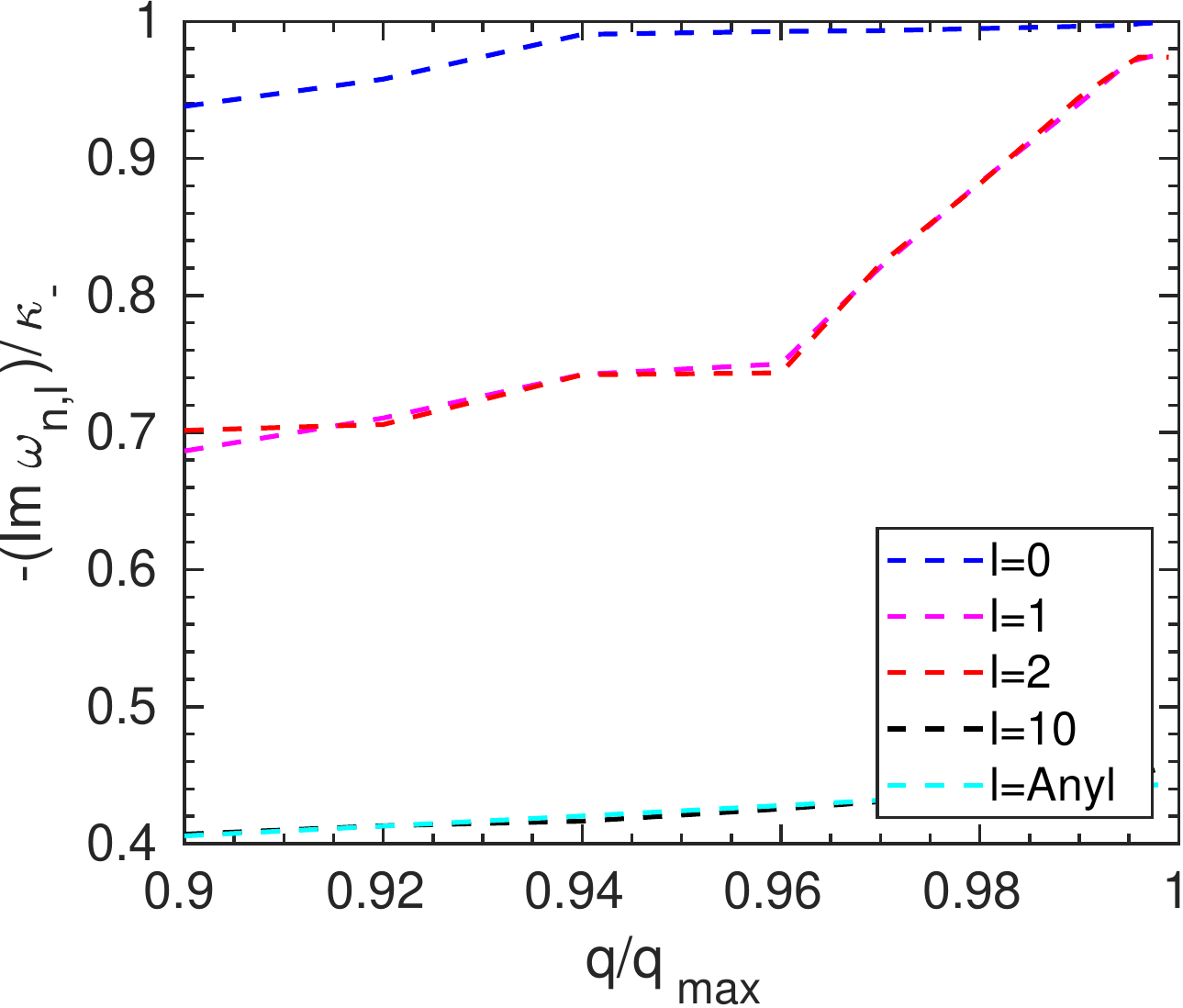}
\caption{We have plotted $\{-(\textrm{Im}~ \omega_{n,l})/ \kappa_{-}\}$ against the ratio $q/q_{\rm max}$, where we have taken the value of the cosmological constant $\Lambda=-0.1$ and the mass $M=1$, for different choices of angular momentum $\ell$.
}
\label{figure3} 
\end{figure*}

\begin{figure*}[h!]
\centering
\includegraphics[scale=0.6]{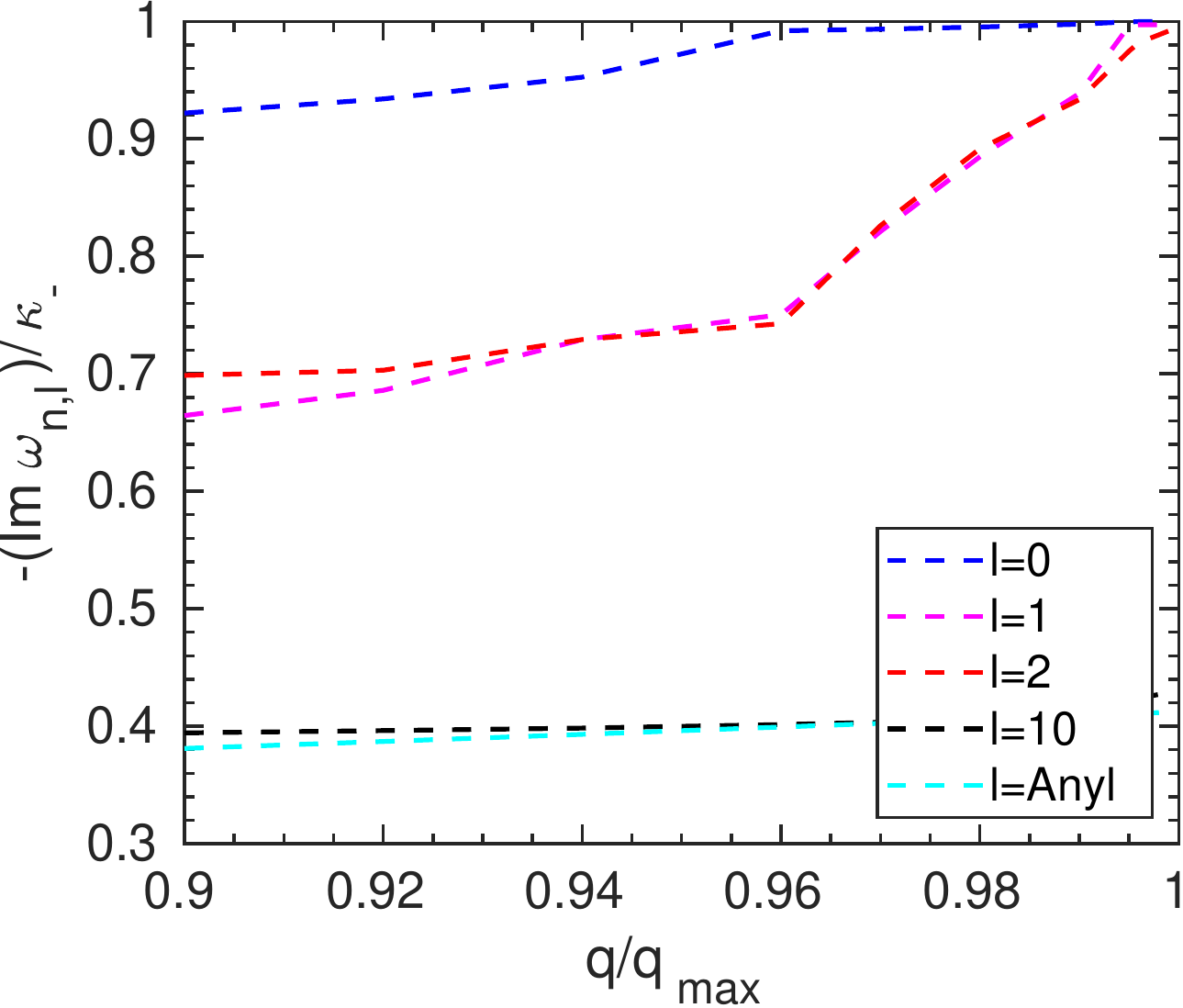}
\caption{We have plotted $\{-(\textrm{Im} \omega_{n,l})/ \kappa_{-}\}$ against $(q/q_{\rm max})$, where we have taken the value of cosmological constant $\Lambda=-0.06$ and mass $M=1$, for different choices of $\ell$.}
\label{figure4} 
\end{figure*}

For a charged BTZ black hole in pure Lovelock gravity of order $N=3$, the metric function $f(r)$ reads, $f(r)=1-\{8M-(q^2/r^4)+\Lambda  r^6\}^{1/3}$, for which the ratio $\{-(\textrm{Im}~\omega_{n,l})/ \kappa_{-}\}$ has been presented in \ref{label3} and \ref{label4}, for different choices of the angular momentum $\ell$ and the cosmological constant $\Lambda$. We have also plotted the ratio $\{-(\textrm{Im}~\omega_{n,l})/ \kappa_{-}\}$ against $(q/q_{\rm max})$, where $q_{\rm max}$ corresponds to the extremal value of the charge, for different choices of angular momentum $\ell$ in \ref{figure3} and \ref{figure4}. We show that the ratio $\{-(\textrm{Im}~\omega_{n,l})/ \kappa_{-}\}$ for the lowest-lying QNM, i.e., $\beta$ is always less than half. Therefore, this ensures that the strong cosmic censorship conjecture is respected for a charged BTZ black hole in pure Lovelock gravity. 
\begin{table*}[h!]          
\centering
\begin{tabular}{cccccccccccccccccccccccccccccccccc}         
\hline\hline                
 $q/q_{max}$ && $\ell=0$ && $\ell=1$ && $\ell=2$ && $\ell=10$  && $\ell=10$ (Analytical)\\ \hline
 0.9 && 0.93798 && 0.686518 && 0.701567 && 0.406967 &&   0.405728\\
 0.92 && 0.957814 && 0.710595 && 0.705895 && 0.413134 &&   0.412883\\
 0.94 && 0.990483 && 0.742405 && 0.742125 && 0.41646 &&   0.420254\\
 0.96 && 0.992679 && 0.750065 && 0.743534 && 0.425379 &&   0.427856\\
 0.97 && 0.993206 && 0.820759 && 0.823032 && 0.430736 && 0.43175  \\
 0.98 && 0.99472 && 0.882366 && 0.881487 && 0.435471 && 0.435709  \\
 0.99 && 0.996369 && 0.940035&& 0.94458 && 0.43928 && 0.439739  \\
 0.995 && 0.997176 && 0.970325 && 0.969527 && 0.445374  && 0.441781
\\ 
 0.996 && 0.998262 && 0.972309 && 0.973914 && 0.450435  && 0.442191\\
 0.999 && 0.999457 && 0.977453 && 0.973897 && 0.456811  && 0.443428\\

\hline\hline                                                    
\end{tabular}                                
\caption{Numerical values of $\{-(\textrm{Im}~ \omega_{n,l})/ \kappa_{-}\}$ have been presented for different choices of angular momentum $\ell$. Here  we have considered the value of the cosmological constant to be $\Lambda=-0.1$ and mass $M=1$, respectively.}
\label{label3}             
\end{table*}  

\begin{table*}[h!]         
\centering 
\begin{tabular}{cccccccccccccccccccccccccccccccccc}         
\hline\hline                
 $q/q_{max}$ && $\ell=0$ && $\ell=1$ && $\ell=2$ && $\ell=10$ && $\ell=10$ (Analytical)\\ 
\hline
 0.9 && 0.921919 && 0.66445 && 0.69861 && 0.394354 && 0.381119  \\
 0.92 && 0.933966 && 0.685912 && 0.70305 && 0.39631 && 0.38704  \\
 0.94 && 0.95246 && 0.729036 && 0.729139 && 0.398436 && 0.393112  \\
 0.96 && 0.992015 && 0.750316 && 0.742653 && 0.401437 && 0.399341  \\
 0.97 && 0.993349 && 0.821108 && 0.826066 && 0.40343 && 0.402517  \\
 0.98 && 0.995164 && 0.884738 && 0.891541 && 0.407975  && 0.405734  \\
 0.99 && 0.997554 && 0.938537 && 0.933365  && 0.417223  && 0.408994  \\
 0.995 && 0.99962 && 0.996126 && 0.974632 && 0.42208 && 0.4106  \\
 0.996 && 0.999729 && 0.99704 && 0.981878 && 0.424151 && 0.410971  \\
 0.999 && 0.999789 && 0.997155 && 0.99183 && 0.428507 && 0.411965  \\
\hline\hline                                                    
\end{tabular}                                
\caption{Numerical values of  $\{-(\textrm{Im}~ \omega_{n,l})/ \kappa_{-}\}$ have been presented for the QNMs with different choices of $\ell$. Here we have considered the value of the cosmological constant $\Lambda=-0.06$ and mass $M=1$.}
\label{label4}             
\end{table*}

As a final remark, we would like to point out that the QNM frequencies and the photon sphere properties are not always in a one-to-one correspondence for black holes in Lovelock gravity. In particular, the scalar, the vector and the tensor parts of the gravitational perturbation for black holes in Lovelock gravity, in the large angular momentum limit, follow an equation identical to \ref{eq13}, but with a potential which is different from that experienced by a photon. This can lead to a significant departure between the QNM frequencies computed using the photon sphere properties \cite{Konoplya:2017wot}. However, for test scalar and test electromagnetic field the correspondence between QNM frequencies and photon sphere again holds true. Since in this work we have considered a test scalar field, the correspondence with the photon sphere will hold good in the present situation. Therefore, the small difference between the analytical and the numerical results as observed from \ref{label3} and \ref{label4}, has possibly originated as we have considered the full effective potential, as in \ref{eq13}, for numerical evaluation of the QNM frequencies, in contrast to the analytical evaluation, which only accounts for the large angular momentum part. It would be interesting to see how the disparity mentioned in \cite{Konoplya:2017wot} works out for gravitational perturbation of charged BTZ black holes in Lovelock gravity. 

\section{Discussion and concluding remarks}\label{conclusion}

The Christodoulou version of the strong cosmic censorship conjecture states that the spacetime cannot be extended beyond the Cauchy horizon with a square-integrable connection. However, the existence of a Cauchy horizon is ubiquitous, any charged and rotating black holes in general relativity and beyond harbours Cauchy horizon and pose a threat to the deterministic nature of gravity. The question then arises whether these spacetimes can be extended beyond the Cauchy horizon in such a way that violation of the Christodoulou condition can be achieved. For asymptotically flat/asymptotically AdS spacetimes the exponential blueshift at the Cauchy horizon makes sure that the strong cosmic censorship conjecture is respected. However, for asymptotically dS black holes with charge this is not the case. A similar scenario can be arrived at for higher dimensional as well as higher curvature theories of gravity. In lower dimensions as well, it has been demonstrated that a rotating BTZ black hole violates strong cosmic censorship conjecture \cite{Dias:2019ery}. More interestingly, the results in the lower dimensional spacetime hold true even when quantum corrections are taken into account \cite{Dias:2019ery}. However, it was pointed out recently in \cite{Emparan:2020rnp}, that the violation of the strong cosmic censorship conjecture occurs for rotating BTZ black hole, only if the first order perturbations are considered, while the violation ceases to exist as higher order terms in the back-reaction, due to the perturbations are included. This is in similar spirit to the result for overcharging/spinning of black holes, where at first order it is possible to overcharge/spin, but at next order, overcharging/spinning becomes impossible \cite{Sorce:2017dst}. 

Following this motivation, in this article, we have explored the fate of the strong cosmic censorship conjecture for a charged BTZ black hole in both general relativity and in higher curvature theories of gravity, namely in pure Lovelock theories. This has been achieved by computing the QNM frequencies for the charged BTZ black hole in general relativity as well as in pure Lovelock theories both analytically as well as numerically. The analytical computation proceeds by determining the Lyapunov exponent, associated with the imaginary part of the QNM frequencies in the eikonal limit. While numerically, we have derived the QNM frequencies for all possible angular momentum values $\ell$. Using these we have derived the lowest lying modes and have identified that the quantity $\beta$ is always less than the critical value $(1/2)$. This ensures that strong cosmic censorship conjecture is respected for charged BTZ black holes, irrespective of whether it is a solution of general relativity or, of pure Lovelock theories. In particular, for charged BTZ solutions, the strong cosmic censorship conjecture holds much stronger in higher order pure Lovelock theories of gravity, in sharp contrast to the four or higher dimensional scenario, in which the presence of higher curvature terms make the violation of the strong cosmic censorship conjecture stronger. In an intuitive way, it is apparent that the stable trapping of null geodesics do not play any significant role in assessing the status of the strong cosmic censorship conjecture, since even without stable trapping, e.g., in the case of Kerr-dS black hole the strong cosmic censorship conjecture is respected. On the other hand, stable trapping of null geodesics play a determining role for asymptotically AdS spacetimes, where the instability of the photon circular orbits results into an violation of the strong cosmic censorship conjecture (e.g., the rotating BTZ black hole spacetime), while for stable trapping of geodesics (e.g., four and higher dimensional asymptotically AdS black holes and the charged BTZ black hole, considered here) the strong cosmic censorship conjecture is respected. This is consistent with the findings of \cite{Emparan:2020rnp}. There are several future prospects of this work, e.g., to derive and study the fate of strong cosmic censorship conjecture for rotating BTZ black holes in pure Lovelock theories of gravity. Moreover, the analysis here is applicable for the massless scalar field, it is to be seen if the same conclusion holds true for the conformal scalar field as well, or, for gravitational perturbation. Finally, any connection between the weak and the strong version of the cosmic censorship conjecture, at least in connection to the AdS black holes will be worth exploring. These we leave for the future.  

\section*{Acknowledgments}

CS thanks the Saha Institute of Nuclear Physics (SINP), Kolkata for financial support. CS also thanks Dipanjan Chakraborty for many useful discussions. Research of SC is funded by the INSPIRE Faculty fellowship from the DST, Government of India (Reg. No. DST/INSPIRE/04/2018/000893) and by the Start-Up Research Grant from SERB, DST, Government of India (Reg. No. SRG/2020/000409). ND warmly acknowledges support of the CAS President's International Fellowship Initiative Grant No. 2020VMA0014. The authors acknowledge the helpful correspondence with Roberto Emparan. 
\bibliography{mastern}

\bibliographystyle{./utphys1}
\end{document}